\documentclass[aps,twocolumn]{revtex4}
\usepackage[colorlinks]{hyperref}
\usepackage[utf8]{inputenc}
\usepackage{mathtools}
\usepackage{epsfig}
\usepackage{subfigure}
\usepackage{graphicx}

\begin{document}
\title{Space-time variation of the $s$ and $c$ quark masses}
\author{V. V. Flambaum and P. Munro-Laylim}
\affiliation{School of Physics, University of New South Wales, Sydney 2052, Australia}

\begin{abstract}
Space-time variation of fundamental physical constants in expanding Universe is predicted by a number of popular models. 
The masses of second generation quarks are larger than first generation quark masses by several orders of magnitude, therefore space-time variation in quark masses may significantly vary between each generation. We evaluate limits on variation in the $s$ and $c$ quark masses from Big Bang nucleosynthesis, Oklo natural nuclear reactor, Yb$^+$, Cs and Rb clock data. 
The construction of a $^{229}$Th nuclear clock is expected to enhance these limits by several orders of magnitude.
Furthermore, constraints are obtained on an oscillating scalar or pseudoscalar cold dark matter field, as interactions of the field with quarks produce variations in quark masses. 
\end{abstract}
	
\maketitle

\section{Introduction}
The discovery of dark matter continues to elude physicists despite it accounting for $85\%$ of the total matter density in the Universe.
Among the range of proposed models for dark matter, axions are one of the most promising candidates. Originally introduced to preserve $CP$-symmetry in QCD \cite{PecceiPRL1977,PecceiPRD1977}, axions and axionlike particles are used to describe light pseudoscalar fields. Light scalar particles with dilaton-like interaction with Standard Model particles are motivated by superstring theory \cite{Veneziano,DAMOUR1994,Damour1994a,DamourVeneziano,DamourVeneziano2} and chameleon models of gravity, see, e.g., Ref.~\cite{Chameleon} and referenced therein. 

Interactions between axions or light scalar dark matter fields with Standard Model particles can lead to the variation of fundamental constants in space-time \cite{Arvanitaki2015,StadnikPRL2015,Stadnik2016,Tilburg2015}. 

There are several new opportunities to search for the variation of fundamental constants, namely the recent launch a network of atomic clocks known as QSNET \cite{Barontini2022} and a new proposed method to use laser-interferometric gravitational-wave detectors \cite{StadnikPRR2019}. Furthermore, as was proposed in Ref.~\cite{Flambaum2006}, the transition frequency between the ground and first excited states of the $^{229}$Th nucleus is highly sensitive to the variation of fundamental constants. A recent review \cite{BeeksNRP2021} outlined the continual advancements towards the construction of a $^{229}$Th nuclear clock, which will hopefully allow us to obtain strong limits on the variation of fundamental constants in the near future. 

Previous works have used various phenomena to constrain variation in the fine structure constant $\alpha$, electron mass $m_e$ and the light quark mass $m_q = (m_u + m_d)/2$ \cite{ShuryakPRD2003,TedescoPRC2006,WiringaPRC2007,WiringaPRC2009,StadnikPRL2015}, however there is a lack of investigation into variation in the $s$ and $c$ quark masses. The masses of the second generation of quarks, $m_s = 93$ MeV and $m_c = 1.3$ GeV, are orders of magnitude bigger than the masses of first generation, $m_q = 3.5$ MeV. Therefore, variation of the second generation quark masses may significantly differ from the first generation. Ref.~\cite{ShuryakPRD2003,TedescoPRC2006} examined variation in the strange quark mass $m_s$ from a variety of phenomena, but their findings are dated and require revision. We provide updated limits on $m_s$ and also present first results on the variation of the charm quark mass, $m_c$.

We place constraints from variation of deuteron binding energy since Big Bang nucleosynthesis, the Oklo natural nuclear reactor, variation in the proton-electron mass ratio, and variation in nuclear magnetic moments. We also determine the constraints expected from a $^{229}$Th nuclear clock. 
  
  Note that  units which are used for the interpretation of the measurements also may vary and this could introduce ambiguity in the interpretation of the results if one considers variation  of the dimensionful parameters - see discussion in Refs. \cite{ShuryakPRD2003,TedescoPRC2006}. Nucleon masses and strong interactions depend mainly on the  QCD scale $\Lambda_{QCD}$ and quark masses.  Therefore, our results may be interpreted as the measurements of the dimensionless parameter $X_q=m_q/\Lambda_{QCD}$ which  does not depend on  units which one uses. It is convenient to assume that QCD scale $\Lambda_{QCD}$ does not vary. We may say that we measure quark masses in units of $\Lambda_{QCD}$.

The variation in quark masses can be due to quark interactions with dark matter. In this work, we consider a non-relativistic scalar or pseudoscalar cold dark matter field, $\phi = \phi_0 \cos(\omega t)$, which oscillates with frequency $\omega = m_\phi c^2 / \hbar$, where $m_\phi$ is mass of the spin-0 dark matter particle. Quadratic-in-$\phi$ interactions between the scalar or pseudoscalar field and Standard Model fermion fields, $f$, are expressed as
\begin{equation}
    \mathcal{L} = \pm \sum_f \frac{\phi^2}{(\Lambda^\prime_f)^2} m_f \bar{f} f,
\end{equation}
where $m_f$ is the fermion mass and $\Lambda^\prime_f$ is a large energy scale that can differ between fermions. Comparison to the Standard Model Lagrangian $ \mathcal{L}=m_f \bar{f}f$ indicates that the fermion masses are altered by the dark matter field
\begin{equation}\label{e:quad_mass_var}
    m_f \rightarrow m_f \left(1 \pm \frac{\phi^2}{(\Lambda^\prime_f)^2}\right).
\end{equation}
Quadratic interactions of $\phi$ mean that there is an oscillating component of $\phi^2$, $\phi_0^2 \cos(2\omega t)/2$, and a non-oscillating component, $\phi_0^2/2$. Therefore, the fundamental constants can experience slow non-oscillating variation, with Ref.~\cite{StadnikPRL2015} obtaining constraints on the field for interactions with photons, electrons, light quarks, and massive vector bosons. Our limits on variation in $m_s$ and $m_c$ are used to constrain the non-oscillating component of dark matter field. Linear-in-$\phi$ interactions of the scalar field with fermions can also be considered to produce only oscillating fermion mass variation
\begin{equation}
    \mathcal{L} = - \sum_f \frac{\phi}{\Lambda_f} m_f \bar{f} f,
\end{equation}
\begin{equation}\label{e:lin_mass_var}
    m_f \rightarrow m_f \left(1 + \frac{\phi}{\Lambda_f}\right).
\end{equation}
We use experimental results from Rb/Cs atomic fountain clocks \cite{HeesPRL2016} to obtain limits on dark matter interactions with second generation quarks. Note that for linear-in-$\phi$ interactions, the field can only be scalar.

\section{Nucleon and Meson Mass Variation}
To evaluate limits on the mass variation of second generation quarks, we first need to obtain the sensitivity in nucleon mass to variation in $m_s$ and $m_c$. In a recent review, Ref.~\cite{FLAG2019} analysed and averaged a range of lattice QCD results, from which we will use the averaged sigma terms
\begin{eqnarray}
    \sigma_s = m_s \langle N|\bar{s}s|N \rangle = &53& {\rm MeV},\\
    \sigma_c = m_c \langle N|\bar{c}c|N \rangle = &78& {\rm MeV},
\end{eqnarray}
where the values are averaged from Ref.~\cite{DurrPRD2012,DurrPRL2016,YangPRD2016,JunnarkarPRD2013,FreemanPRD2013} for $\sigma_s$ and Ref.~\cite{JunnarkarPRD2013,BaliPRD2016,AbdelRehimPRL2016,GongPRD2013,FreemanPRD2013} for $\sigma_c$. Therefore, the nucleon mass sensitivity to each of the second generation quark masses from the strange and charm seas are found to be
\begin{eqnarray}
    \frac{\delta m_N}{m_N} = \frac{\sigma_s}{m_N} \frac{\delta m_s}{m_s} = 0.056 \frac{\delta m_s}{m_s}, \label{eq:delm_N1}\\
    \frac{\delta m_N}{m_N} = \frac{\sigma_c}{m_N} \frac{\delta m_c}{m_c} = 0.083 \frac{\delta m_c}{m_c}, \label{eq:delm_N2}
\end{eqnarray}
for nucleon mass $m_N = 939$ MeV. 
Our limits on second generation quark mass variation from deuteron binding energy variation and the Oklo nuclear reactor also require us to account for variation in the strong nuclear potential. The Walecka model \cite{Walecka} is used to express the strong nuclear potential via the exchange of $\sigma$ and $\omega$ mesons. 
We evaluate the sensitivities of meson masses to quark mass variation (see Appendix~\ref{app:meson})
\begin{eqnarray}
    \frac{\delta m_\sigma}{m_\sigma} &=& 0.33 \frac{\delta m_s}{m_s},\\
    \frac{\delta m_\sigma}{m_\sigma} &=& 0.10 \frac{\delta m_c}{m_c},\\
    \frac{\delta m_\omega}{m_\omega} &=& 0.045 \frac{\delta m_s}{m_s},\\
    \frac{\delta m_\omega}{m_\omega} &=& 0.067 \frac{\delta m_c}{m_c}.
\end{eqnarray}

\section{Variation in second generation quark masses}
\subsection{Big Bang Nucleosynthesis}
In Ref.~\cite{ShuryakPRD2003}, variation in deuteron binding energy was estimated using the Walecka model
\begin{equation}\label{e:deut_sens}
    \frac{\delta Q_d}{Q_d} = -48 \frac{\delta m_\sigma}{m_\sigma} + 50 \frac{\delta m_\omega}{m_\omega} + 6 \frac{\delta m_N}{m_N}.
\end{equation}
Note that the Walecka model cannot be used to correctly describe deuteron binding energy variation, as the tensor forces (from $\pi$ and $\rho$ exchange) are required to account for all spin-dependent forces. Instead of accounting for the tensor forces directly, the authors of Ref.~\cite{ShuryakPRD2003} modified the Walecka model by reducing $g_v^2$ by a factor of 0.953 to obtain Eq.~\ref{e:deut_sens}. This estimate allowed them to correctly evaluate the deuteron binding energy. Therefore, using our results for nucleon and meson mass variation, we find the sensitivity in deuteron binding energy due to second generation quark mass variation
\begin{eqnarray}
    \frac{\delta Q_d}{Q_d} &=& -13 \frac{\delta m_s}{m_s},\\
    \frac{\delta Q_d}{Q_d} &=& -0.95 \frac{\delta m_c}{m_c}.
\end{eqnarray}
From Ref.~\cite{DmitrievPRD2004}, the limit on the variation of deuteron binding energy from Big Bang nucleosynthesis to today is $\delta Q_d / Q_d = -0.019\pm0.005$. Therefore, we obtain limits on the mass variation of second generation quarks since Big Bang nucleosynthesis
\begin{eqnarray}
    \left| \frac{\delta m_s}{m_s} \right| &<& 0.0018, \label{e:bbn_s}\\
    \left| \frac{\delta m_c}{m_c} \right| &<& 0.025. \label{e:bbn_c}
\end{eqnarray}

\subsection{Oklo natural nuclear reactor}
The Oklo natural nuclear reactor is a self-sustaining nuclear fission reactor that has been active for around 2 billion years. 
At Oklo, the disappearance of isotopes with near-zero neutron resonance energy, most notably $^{149}$Sm, can be used to constrain the shift in the lowest resonance, $\delta E$.
The strongest limit on this energy shift is $|\delta E| < 0.02$ eV \cite{Damour1996}. From Ref.~\cite{ShuryakPRD2003}, the variation of the resonance position is given by
\begin{eqnarray}
    \delta E = V_0 \left( 8.6 \frac{\delta m_\sigma}{m_\sigma} - 6.6 \frac{\delta m_\omega}{m_\omega} - \frac{\delta m_N}{m_N}\right),
\end{eqnarray}
from which we obtain the variation of resonance position due to $s$ and $c$ quark mass variation, with $V_0 \approx 50$ MeV,
\begin{eqnarray}
    \delta E &=& 120\; {\rm MeV} \times \frac{\delta m_s}{m_s},\\
    \delta E &=& 17\; {\rm MeV} \times \frac{\delta m_c}{m_c}.
\end{eqnarray}
Therefore, we constrain the mass variation of second generation quarks
\begin{eqnarray}
    \left| \frac{\delta m_s}{m_s} \right| &<& 1.7\times10^{-10},\\
    \left| \frac{\delta m_c}{m_c} \right| &<& 1.2\times10^{-9}.
\end{eqnarray}
Assuming a constant rate of variation over $1.8\times10^9$ years, these limits correspond to the constrained rate of variation
\begin{eqnarray}
    \left| \frac{1}{m_s}\frac{dm_s}{dt} \right| &<& 9.4\times10^{-20}\; {\rm yr}^{-1},\\
    \left| \frac{1}{m_c}\frac{dm_c}{dt} \right| &<& 6.7\times10^{-19}\; {\rm yr}^{-1}.
\end{eqnarray}

\subsection{Proton-electron mass ratio}
Recent works have measured bounds on the time variation of the proton-electron mass ratio $\mu_{pe}=m_p/m_e$ with improving accuracy \cite{LangePRL2021,McGrew2019}. We assume that $m_e$ does not vary and obtain limits on second generation quark mass variation, and therefore $\delta \mu_{pe}/\mu_{pe} = \delta m_p/m_p$ with the nucleon mass in units of electron mass. 
Using our previously mentioned values in Eq.~\ref{eq:delm_N1} and \ref{eq:delm_N2}, we get the limits from proton-electron mass variation
\begin{eqnarray}
    \frac{\delta \mu_{pe}}{\mu_{pe}} &\approx& 0.056 \frac{\delta m_s}{m_s},\\
    \frac{\delta \mu_{pe}}{\mu_{pe}} &\approx& 0.083 \frac{\delta m_c}{m_c}.
\end{eqnarray}
Ref.~\cite{LangePRL2021} obtained the strong limit of $(1/\mu_{pe})(d\mu_{pe}/dt) = -8(36)\times10^{-18}$ yr$^{-1}$ from comparing transition frequencies in optical Yb$^+$ clocks and Cs clock.
We use this result to obtain our limits on the second generation quark mass variation
\begin{eqnarray}
    \left| \frac{1}{m_s}\frac{dm_s}{dt} \right| &<& 7.9\times10^{-16}\; {\rm yr}^{-1},\\
    \left| \frac{1}{m_c}\frac{dm_c}{dt} \right| &<& 5.3\times10^{-16}\; {\rm yr}^{-1}.
\end{eqnarray}

\subsection{Nuclear magnetic moments}
It was first noted in Ref.~\cite{Karshenboim2000} that the ratios of hyperfine structure intervals between different atoms are sensitive to the variation of nuclear magnetic moments. 
In Ref.~\cite{LeinweberPRD2004}, chiral perturbation theory was used to calculate nucleon magnetic moment variation, $\mu_p$ and $\mu_n$ for protons and neutrons respectively, from the $s$ quark. These results were then implemented in Ref.~\cite{TedescoPRC2006} to calculate variation of the dimensionless nuclear magnetic moment (in units of nuclear magneton), $\mu$, for a variety of nuclei of the form
\begin{equation}
    \frac{\delta \mu}{\mu} = \kappa_s \frac{\delta m_s}{m_s},
\end{equation}
where values for $\kappa_s$ are nucleus-dependent and are contained in Ref.~\cite{TedescoPRC2006}. The $c$ quark is too heavy for chiral perturbation theory, and the corresponding contribution to variation in nuclear magnetic moments is likely small. Let us define the parameter
\begin{equation}
    V \equiv \frac{A}{E} = {\rm const} \times [\alpha^2 F_{\mathrm rel} (Z\alpha)] \left(\mu\frac{m_e}{m_p}\right),
\end{equation}
where $A$ is the hyperfine structure constant and $E = m_e e^4/\hbar^2$ is the atomic unit of energy. The first set of brackets relate the dependence on $\alpha$ and the relativistic correction factor (Casimir factor), $F_{\mathrm rel}$. The last set of brackets contain the dimensionless nuclear magnetic moment, $\mu$ (nuclear magnetic moment $M = \mu (e\hbar/2m_pc)$), and the proton and electron masses, $m_p$ and $m_e$. Assuming that only the mass of second generation quarks varies, only $\mu$ and $m_p$ contribute to the variation of $V$
\begin{equation}
    \frac{\delta V}{V} = \frac{\delta \mu}{\mu} - \frac{\delta m_p}{m_p}.
\end{equation}
Therefore, we consider the ratio of hyperfine structure constants, $X(a_1/a_2) = V(a_1)/V(a_2)$, between two atoms, $a_1$ and $a_2$. The variation in this ratio from variation in $m_s$ is then expressed as
\begin{eqnarray}
    \frac{\delta X(a_1/a_2)}{X(a_1/a_2)} = (\kappa_{s,a_1} - \kappa_{s,a_2}) \frac{\delta m_s}{m_s}.\label{e:delX}
\end{eqnarray}
Limits on slow-drift variation between $^{87}$Rb and $^{133}$Cs hyperfine transitions from the dual atomic fountain clock FO2 at LNE-SYRTE in Ref. \cite{Abgrall2015} are $\left(1/X\left({\rm Rb/Cs}\right)\right)\left(dX\left({\rm Rb/Cs}\right)/dt\right) < -11.6(6.1)\times10^{-17}$ yr$^{-1}$ from numerous measurements spanning over a decade. Therefore, we obtain the limit on variation in $s$ quark mass
\begin{equation}
    \left|\frac{1}{m_s}\frac{dm_s}{dt}\right| < 9.8\times10^{-15}\; {\rm yr}^{-1},
\end{equation}
where we have used $\kappa_{s,\rm Rb} = -0.010$ and $\kappa_{s,\rm Cs} = 0.008$~\cite{TedescoPRC2006}.

\subsection{Thorium nuclear clock}
Ref.~\cite{Flambaum2006} noted that a $^{229}$Th nuclear clock transition between the ground and first excited states will be highly sensitive to variations in quark masses. Later work was done to calculate the difference in Coulomb and kinetic energies between the $3/2^+$ and $5/2^+$ states using Hartree-Fock (HF) and Hartree-Fock-Bogoliubov (HFB) calculations \cite{LitvinovaPRC2009}. The difference in Coulomb energies were used to obtain the transition frequency sensitivity to variation in $\alpha$, while  the difference in kinetic energies $T=p^2/2m$ can be used to obtain sensitivity to variation in nucleon masses. 
%
For a transition energy, $\omega = E_1 - E_0$ 
the sensitivity  to variation in nucleon mass may be estimated as 
\begin{eqnarray}
    \delta \omega 
    &=& -(\Delta T_p + \Delta T_n) \frac{\delta m_N}{m_N},
\end{eqnarray}
We assume $m_p = m_n = m_N$. In Ref.~\cite{LitvinovaPRC2009}, values for $\Delta T_p$ and $\Delta T_n$ were obtained using HF and HFB calculations. 
The HFB results for the energy functional SIII \cite{SIII} are used due to their similarity to the experimental energy of the $^{229}$Th 5/2$^+$ ground state, as well as close similarity to semi-empirical estimates on Coulomb, kinetic and strong energy shifts from Ref.~\cite{FadeevPRC2022}. We obtain
\begin{equation}
    \delta \omega = -65\;{\rm keV}\times \frac{\delta m_N}{m_N}.
\end{equation}\\
As was shown in Ref.~\cite{CampbellPRL2012}, a $^{229}$Th nuclear clock can reach precision $\delta \omega / \omega \sim 10^{-19}$. For the transition frequency between the 3/2$^+$ and 5/2$^+$ states, $\omega \approx  8$ eV \cite{SikorskyPRL2020}, we expect limits on variation in the transition frequency from $^{229}$Th clocks to reach $\delta \omega \approx 8\times10^{-19}$~eV~yr$^{-1}$. For this variation be due from variation in the second generation quark masses, we obtain the expected limits attainable by a $^{229}$Th clock
\begin{eqnarray}
    \left| \frac{1}{m_s}\frac{dm_s}{dt} \right| &\lesssim& 2\times10^{-22}\; {\rm yr}^{-1},\\
    \left| \frac{1}{m_c}\frac{dm_c}{dt} \right| &\lesssim& 1\times10^{-22}\; {\rm yr}^{-1}.
\end{eqnarray}
These expected constraints are the strongest out of all phenomena we have investigated; 
the construction of a $^{229}$Th clock is highly anticipated for investigations into the variation of fundamental constants.

\section{Limits on scalar dark matter}
\subsection{Limits from non-oscillating contribution}
\begin{figure}[tb]\centering
    \subfigure{\includegraphics[width=0.45\textwidth]{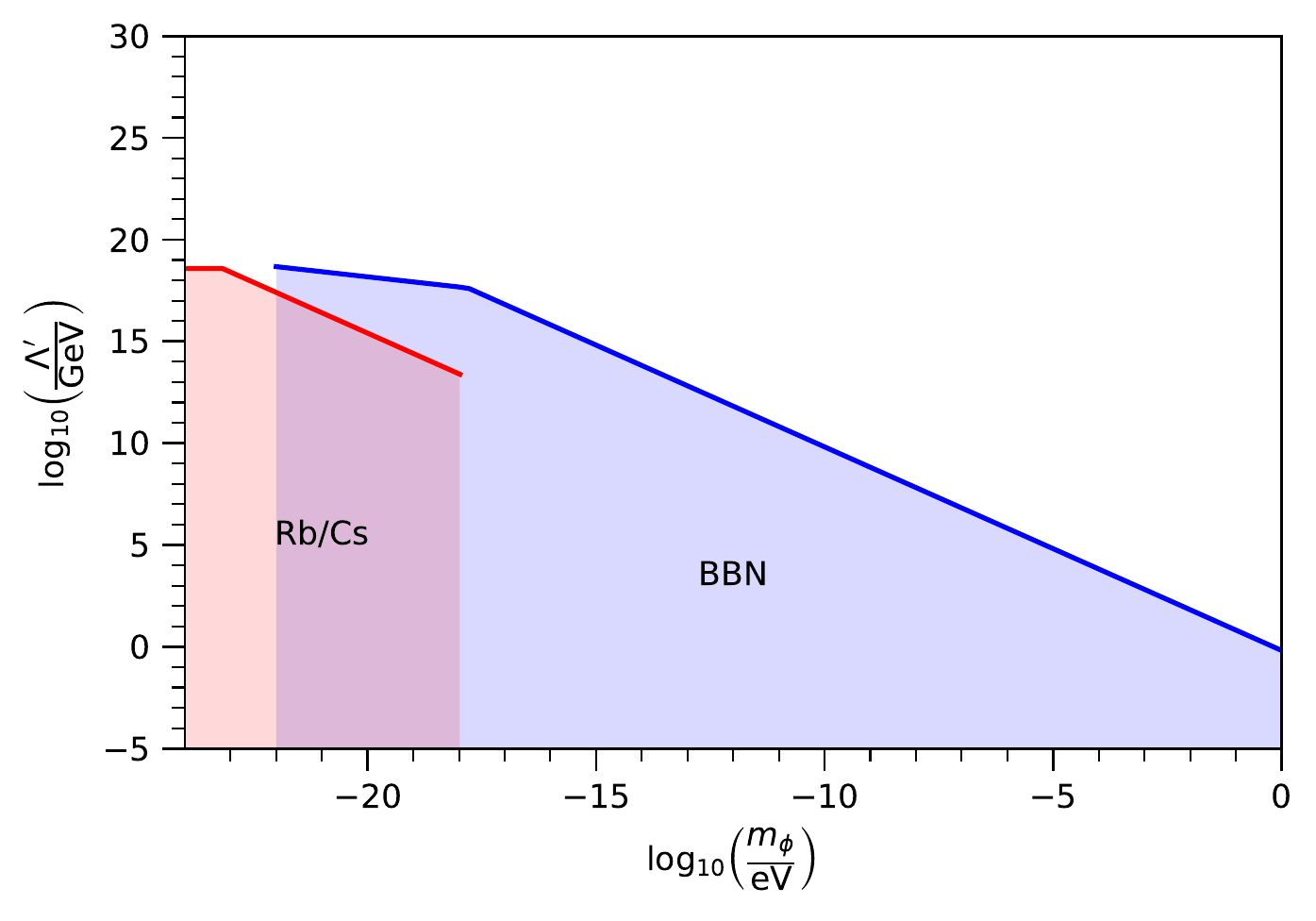}}
    \subfigure{\includegraphics[width=0.45\textwidth]{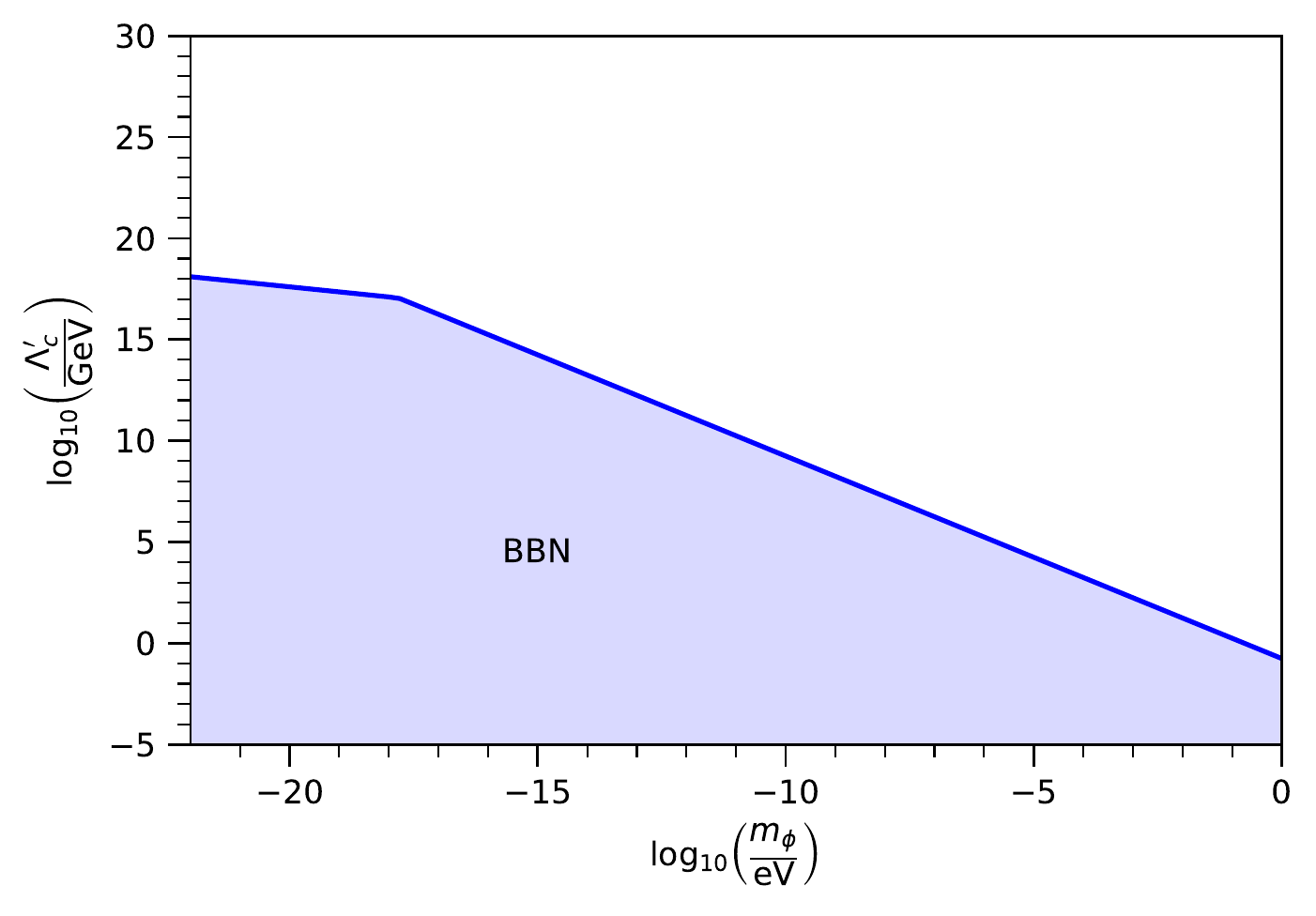}}
    \caption{\label{fig:dm_lim} From top to bottom: Limits on quadratic interaction between scalar and pseudoscalar dark matter field $\phi$ and $s$ and $c$ quarks. Shaded region represents excluded values. Blue region corresponds to limits from variation in deuteron binding energy since Big Bang nucleosynthesis \cite{DmitrievPRD2004}. 
      Red region corresponds to  limits based on measurement of oscillating effect in  Ref.~\cite{HeesPRL2016}.}
\end{figure}

We first consider quadratic interactions of an oscillating scalar or pseudoscalar dark matter field, $\phi = \phi_0 \cos(\omega t)$, with $s$ and $c$ quarks. Using Eq.~\ref{e:quad_mass_var}, the fractional variation in fermion mass from the field
\begin{equation}\label{e:quad_frac_m}
    \bigg|\frac{\delta m_f}{m_f}\bigg| = \frac{\phi^2}{(\Lambda^\prime_f)^2}.
\end{equation}

This variation has a non-oscillating component, $\phi_0^2/2(\Lambda^\prime_f)^2$, which induces non-oscillating variation of fundamental constants from changes in the dark matter density by an oscillating field $\rho = m_\phi^2 \langle \phi^2 \rangle$, where $\rho_{DM,0}=1.3\times10^{-6}\; {\rm GeV\; cm}^{-3}$ is average dark matter density \cite{RPP2022}. 
Note that there is a lower limit $m_\phi > 10^{-22}$ eV on the field to allow correct large scale structure formation in galaxies \cite{Bozek2015,Schive2016,Armengaud2017}; the field can only contribute to a fraction of the dark matter density below this limit.

We consider limits on $\Lambda^\prime_f$ and $m_\phi$ for two cases: $m_\phi > H$ and $m_\phi < H$, where $H$ is the Hubble parameter during Big Bang nucleosynthesis. In the former case, $\phi$ remains an oscillating field, and in the latter case, the field freezes due to Hubble friction, becoming a non-oscillating constant field. For an early radiation-dominated universe, the Hubble parameter scales as $H=1/2t$. Therefore, at the time of Big Bang nucleosynthesis, 
we have the cross-over value $m_\phi \sim 10^{-16}$ eV.

Using Eq.~\ref{e:quad_frac_m} and our limits on variation in $m_s$ and $m_c$ from Eq.~\ref{e:bbn_s} and \ref{e:bbn_c}, we evaluate constraints for $\Lambda^\prime_f$ and $m_\phi$ (see Appendix  for details) for $m_\phi > H$
\begin{eqnarray}
    (\Lambda^\prime_s)^2 > \frac{4.3\times10^{17} \; {\rm eV}^{4}}{m_\phi^2},\\
    (\Lambda^\prime_c)^2 > \frac{3.1\times10^{16} \; {\rm eV}^{4}}{m_\phi^2},
\end{eqnarray}
and for $m_\phi < H$
\begin{eqnarray}
    (\Lambda^\prime_s)^2 > \frac{2.2\times10^{44} \; {\rm eV}^{5/2}}{m_\phi^{1/2}},\\
    (\Lambda^\prime_c)^2 > \frac{1.6\times10^{43} \; {\rm eV}^{5/2}}{m_\phi^{1/2}}.
\end{eqnarray}
Our results are presented in Fig.~\ref{fig:dm_lim}, where the change in gradient illustrates the cross-over value between $m_\phi > H$ and $m_\phi < H$.

\begin{figure}[tb]\centering
    \subfigure{\includegraphics[width=0.45\textwidth]{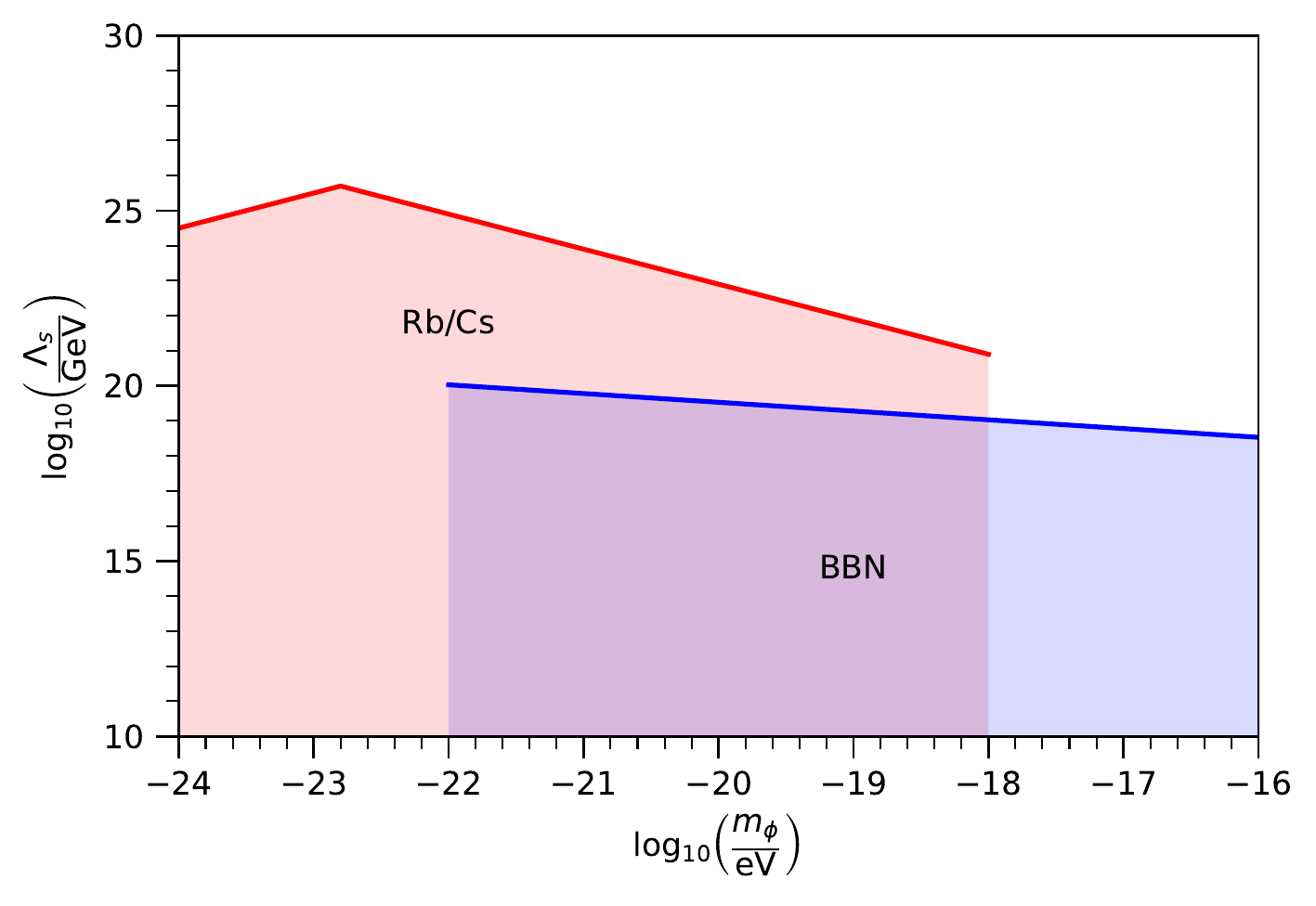}}
    \subfigure{\includegraphics[width=0.45\textwidth]{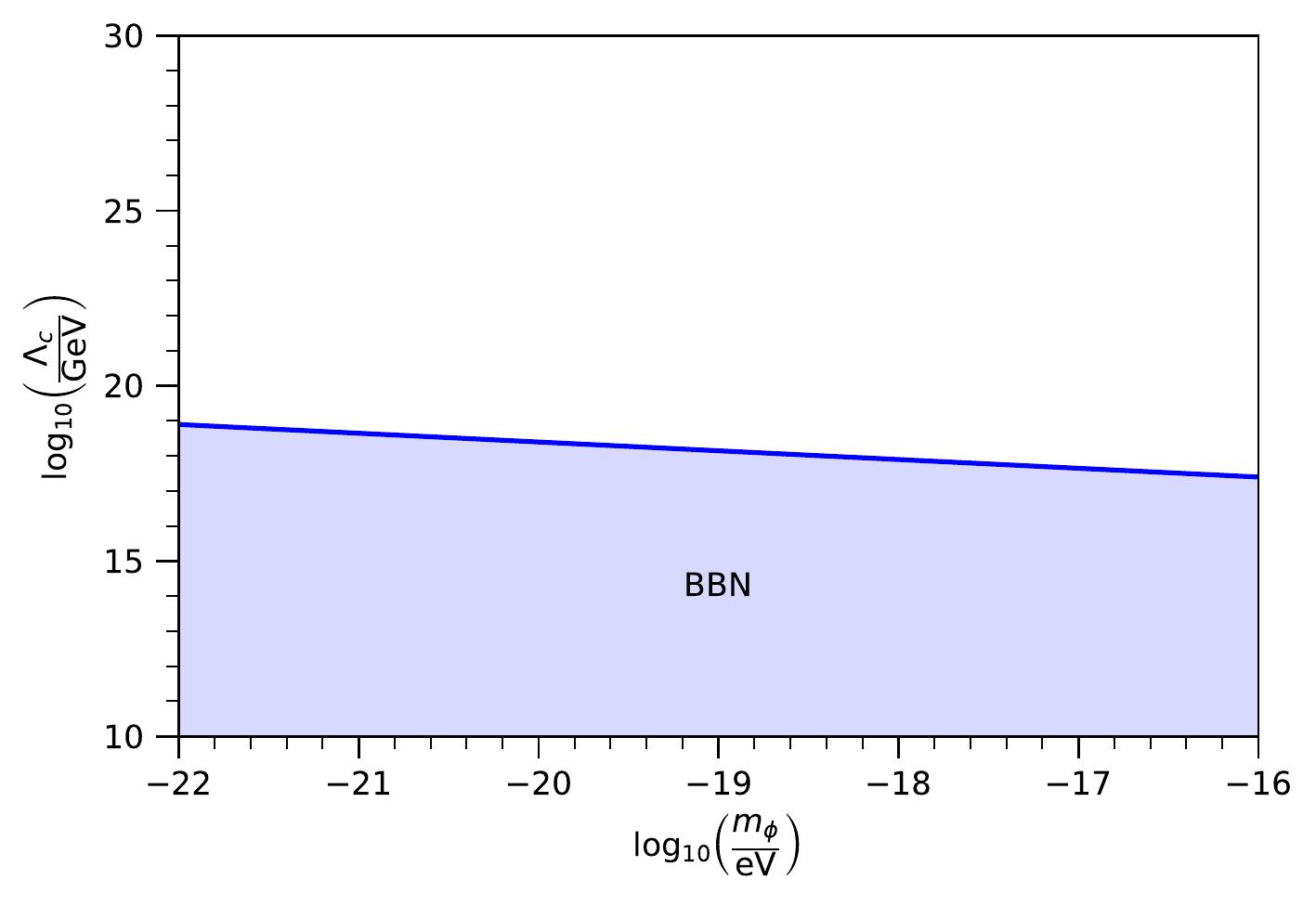}}
    \caption{\label{fig:dm_lim_lin} From top to bottom: Limits on linear interaction between scalar dark matter field $\phi$ and $s$ and $c$ quarks. Shaded region represents excluded values. Blue region corresponds to limits from variation in deuteron binding energy since Big Bang nucleosynthesis \cite{DmitrievPRD2004}. 
    Red region corresponds to  limits based on measurement of oscillating effect in Ref.~\cite{HeesPRL2016}.}
\end{figure}

We also consider linear-in-$\phi$ interactions between an oscillating scalar dark matter field with second generation quarks. From Eq.~\ref{e:lin_mass_var}, the fractional variation in mass oscillates with the field
\begin{equation}
    \bigg|\frac{\delta m_f}{m_f}\bigg| = \frac{\phi}{\Lambda_f}.
\end{equation}
However, the field freezes when $m_\phi < H$, inducing slow non-oscillating variation in fundamental constants. Therefore, similarly to the quadratic interaction, we obtain the constraints, also presented in Fig.~\ref{fig:dm_lim_lin},
\begin{eqnarray}
    \Lambda_s > \frac{3.4\times10^{23} \; {\rm eV}^{5/4}}{m_\phi^{1/4}},\\
    \Lambda_c > \frac{2.5\times10^{22} \; {\rm eV}^{5/4}}{m_\phi^{1/4}}.
\end{eqnarray}

\subsection{Limits from oscillating contribution}
In Ref.~\cite{HeesPRL2016}, the dual rubidium and cesium atomic fountain clock FO2 at LNE-SYRTE was used to search for oscillations in the ratio of Rb/Cs hyperfine transition frequencies. As previously demonstrated in Eq.~\ref{e:delX}, the ratio of hyperfine transition frequencies is sensitive to variations in $m_s$. Since the oscillating mass variation can originate from interactions with $\phi$, oscillations in this ratio can be used to constrain the interactions.

The measurements for limits on the amplitude of oscillations from Ref.~\cite{HeesPRL2016} are used to find limits for the linear and quadratic interactions. Our constraints are presented in Fig.~\ref{fig:dm_lim_lin} Note that we use $\phi_0 = \sqrt{\rho_{DM}}/m_\phi$ for the local dark matter density $\rho_{DM} = 0.3$ GeV cm$^{-3}$.

\section{Conclusion}
In this work, we examined variation in strange and charm quark masses. We obtained our strongest limits on the rate of mass variation, $(1/m)(dm/dt) \lesssim 10^{-20}$ yr$^{-1}$, using measurements from the Oklo natural nuclear reactor in Ref.~\cite{Damour1996}. 
Assuming an expected accuracy of $\delta\omega \sim 10^{-19}$ \cite{CampbellPRL2012}, we show that a $^{229}$Th nuclear clock is expected to provide the best limits of $(1/m)(dm/dt) \lesssim 10^{-22}$ yr$^{-1}$. 

An oscillating scalar or pseudoscalar cold dark matter field that interacts with $s$ and $c$ quarks can produce variations in quark masses. We place limits in Fig.~\ref{fig:dm_lim} on previously unconstrained quadratic-in-$\phi$ interaction parameters, $\Lambda^\prime_s$ and $\Lambda^\prime_c$, and mass of a spin-0 scalar or pseudoscalar dark matter particle, $m_\phi$. Linear-in-$\phi$ interactions between the dark matter field and second generation quarks are also examined, where our limits on $\Lambda_s$, $\Lambda_c$ and $m_\phi$ are presented in Fig.~\ref{fig:dm_lim_lin}.

\section*{Acknowledgements}
This work was supported by the Australian Research Council Grants No. DP190100974 and DP200100150.

\appendix
\section{Meson Mass Variation}\label{app:meson}
 We use the Walecka model \cite{Walecka} to express the strong nuclear potential via the exchange of $\sigma$ and $\omega$ mesons
\begin{equation}
    V = -\frac{g_s^2}{4\pi}\frac{e^{-m_{\sigma}r}}{r} + \frac{g_v^2}{4\pi}\frac{e^{-m_{\omega}r}}{r}.
\end{equation}
For masses $m_\sigma \approx 500$ MeV and $m_\omega \approx 780$ MeV, the coupling strengths are $g_s^2 \approx 100$ and $g_v^2 \approx 190$ \cite{Serot1984}. 
In the simple constituent quark picture, we estimate meson mass sensitivity to variation in quark mass from the quark sea to be two-thirds of the nucleon sensitivity.
This leads to our results
\begin{eqnarray}
    \frac{\delta m_\sigma}{m_\sigma} &=& 0.071 \frac{\delta m_s}{m_s},\\
    \frac{\delta m_\sigma}{m_\sigma} &=& 0.10 \frac{\delta m_c}{m_c},\\
    \frac{\delta m_\omega}{m_\omega} &=& 0.045 \frac{\delta m_s}{m_s},\\
    \frac{\delta m_\omega}{m_\omega} &=& 0.067 \frac{\delta m_c}{m_c},
\end{eqnarray}
where we use the masses $m_\sigma \approx 500$ MeV and $m_\omega \approx 780$ MeV \cite{RPP2022}.

For the $\sigma$ meson, it was noted in Ref.~\cite{ShuryakPRD2003} that there are additional contributions to the variation from the $s$ quark. The $\sigma$ meson can be approximated as the $SU(3)$ singlet state $\sigma = (\bar{u}u+\bar{d}d+\bar{s}s)/\sqrt{3}$, leading to the valence contribution
\begin{equation}
    \frac{\delta m_\sigma}{m_\sigma} = \frac{m_s \langle\sigma|\bar{s}s|\sigma\rangle}{m_\sigma} \frac{\delta m_s}{m_s} = 0.12 \frac{\delta m_s}{m_s}.
\end{equation}
Furthermore, there are additional contributions from mixing with virtual $\bar{K}K$ and $\eta\eta$ pairs (see Ref.~\cite{ShuryakPRD2003} for details)
\begin{equation}
    \frac{\delta m_\sigma}{m_\sigma} = 0.14 \frac{\delta m_s}{m_s}.
\end{equation}
Therefore, the total sensitivity in the $\sigma$ meson mass from the $s$ quark mass is given by the sum of all contributions
\begin{equation}
    \frac{\delta m_\sigma}{m_\sigma} = 0.33 \frac{\delta m_s}{m_s}.
\end{equation}

\section{Big Bang Nucleosynthesis Limits}\label{app:bbn}
We first consider $m_\phi > H$ and use the relation for the dark matter density
\begin{equation}\label{e:dm_density}
    \rho(z) = \rho_{DM,0} (1+z)^3,
\end{equation}
where $\rho_{DM,0}=1.3\times10^{-6}\; {\rm GeV\; cm}^{-3}$ is average dark matter density \cite{RPP2022} and $z$ is the redshift parameter. The redshift parameter for deuterium formation during Big Bang nucleosynthesis is $z=4.3\times10^{8}$, therefore the variation in the quark mass is
\begin{equation}
    \bigg|\frac{\delta m_f}{m_f}\bigg| = \frac{7.8\times10^{14} \; {\rm eV}^{4}}{m_\phi^2 (\Lambda^\prime_f)^2}.
\end{equation}
Using our limits on variation in $m_s$ and $m_c$ from variation in deuteron binding energy since Bang Bang nucleosynthesis, the resulting constraints are
\begin{eqnarray}
    (\Lambda^\prime_s)^2 > \frac{4.3\times10^{17} \; {\rm eV}^{4}}{m_\phi^2},\\
    (\Lambda^\prime_c)^2 > \frac{3.1\times10^{16} \; {\rm eV}^{4}}{m_\phi^2}.
\end{eqnarray}

We now consider the case where $m_\phi < H$. As $z$ is very large, we assume a radiation-dominated universe and $H = H_0 \Omega_{r,0}^{1/2}(1+z)^2$ for $H_0 = 67$ km s$^{-1}$ Mpc$^{-1}$ and radiation density $\Omega_{r,0} = 9.26\times10^{-5}$ \cite{RPP2022}.
Therefore, we use Eq.~\ref{e:dm_density} to get
\begin{eqnarray}
    \rho 
    = \frac{\rho_{DM,0}}{\Omega_{r,0}^{3/4}} \bigg(\frac{m_\phi}{H_0}\bigg)^{3/2}.
\end{eqnarray}
The energy density for a non-oscillating field is $\rho = m_\phi^2 \langle\phi^2\rangle/2$, thus we obtain the variation in quark mass 
\begin{eqnarray}
    \bigg|\frac{\delta m_f}{m_f}\bigg| 
    = \frac{3.9\times10^{41}\; {\rm eV}^{5/2}}{m_\phi^{1/2} (\Lambda^\prime_f)^2}.
\end{eqnarray}
Using our limits on variation in $m_s$ and $m_c$, we find the constraints
\begin{eqnarray}
    (\Lambda^\prime_s)^2 > \frac{2.2\times10^{44} \; {\rm eV}^{5/2}}{m_\phi^{1/2}},\\
    (\Lambda^\prime_c)^2 > \frac{1.6\times10^{43} \; {\rm eV}^{5/2}}{m_\phi^{1/2}}.
\end{eqnarray}

\end{document}